\documentclass[twocolumn,showpacs,amsmath,amssymb]{revtex4-1}

\usepackage{setspace}
\usepackage{graphicx}
\usepackage{amsmath}
\usepackage{lipsum}
\usepackage{cases}
\usepackage{longtable}
\usepackage{afterpage}
\usepackage{rotating}
\usepackage{threeparttable}
\usepackage{booktabs}
\usepackage{multirow}
\usepackage{subfigure}
\usepackage{array}
\usepackage{color}
\usepackage{float}
\usepackage[section]{placeins}
\usepackage{longtable}

\usepackage{bm}
\usepackage{lipsum}
\usepackage{appendix}
\usepackage{soul}
\usepackage[colorlinks=true,linkcolor=red,citecolor=blue,urlcolor=red,]{hyperref}

\begin{document}

\renewcommand\arraystretch{1}
\title{Relativistic coupled-cluster-theory analysis of the hyperfine interaction of Ra$^{+}$ isotopes}
\author{Fei-Chen Li,$^{1,3}$ Yong-Bo Tang,$^{2,*}$ Hao-Xue Qiao,$^{1,*}$ and Ting-Yun Shi$^{3}$ }
\affiliation {$^1$Department of Physics, Wuhan University, Wuhan, 430072, China}
\affiliation {$^2$College of Engineering Physics, Shenzhen Technology University, Shenzhen, 518118, China}
\affiliation {$^3$State Key Laboratory of Magnetic Resonance and Atomic and Molecular Physics, Wuhan
Institute of Physics and Mathematics, Innovation Academy for Precision Measurement Science and Technology, Chinese Academy of Sciences, Wuhan, 430071, China}
\email{tangyongbo@sztu.edu.cn, qhx@whu.edu.cn}
\date{\today}

\begin{abstract}
Hyperfine-structure constants of odd Ra$^{+}$ due to the interactions of nuclear magnetic dipole, electric quadrupole, and magnetic octupole moments with the electrons are investigated in the framework of relativistic coupled-cluster method within single- and double-excitation approximation. The calculated energies and magnetic dipole hyperfine-structure constants $A$ exhibit a good agreement with available experimental values. Combining with the experimental electric quadrupole hyperfine-structure constant, we also extracted the electric quadrupole moments $Q$ of $^{209,211,221,223}$Ra. Our $Q$($^{221}$Ra) and $Q$($^{223}$Ra) are consistent with the referenced values from a semi-empirical analysis (Z. Phys. D: At., Mol. Clusters 11, 105 (1988)), but $Q(^{211}$Ra)=$0.33(2)$ is smaller than the referenced value $0.48(4)$ by about 30\%. Furthermore, we also performed a procedure for assessing the contributions of  magnetic octupole moment to the hyperfine splitting. The sensitivity of hyperfine-structure interval measurements in $^{223}$Ra$^{+}$ that can reveal the effect caused by the nuclear octupole moment are found to be on the order of kHz.

\end{abstract}

\pacs{ } \maketitle
\section{INTRODUCTION}
Hyperfine-structure(HFS) of the atomic energy level is caused by the interaction between electrons and the electromagnetic multipole moments of the nucleus. Theoretical studies of HFS are of great significance in precisely determining different atomic properties and obtaining nuclear structure information. For example, the interpretation of the atomic parity-nonconservation (PNC) and electric dipole moment (EDM) experiments need to theoretically provided the weak matrix elements of PNC amplitude and the P,T-odd interaction constants of EDM which cannot be directly measured~\cite{Nataraj2008prl,Dzuba2012prl,Porsev2012prl}. The accuracy of these calculations can be reliably estimated by comparing the experimental and theoretical values of the magnetic dipole HFS constant, because all of them are sensitive to the wave function behavior near the nucleus, although the three resulting mechanisms completely different. With the advancement of spectroscopy technologies, the hyperfine splittings can be measured very precisely. The comparison between the HFS spectrum and the corresponding high-precision calculation can offer an alternative route to determine the electromagnetic multipole moments of the nucleus in a nuclear model-independent way. This is also currently one of the most accurate methods for determining the nuclear moment, especially the nuclear quadrupole moments $Q$ and octupole moments $\Omega$ of heavy nuclei.

Ra$^{+}$, as the heaviest alkaline earth metal ion in the periodic table, has been proposed as a promising candidate for the measurement of atomic parity-nonconservation (PNC) effect~\cite{Geetha1998pra,Ginges2004pr,Dzuba2001pra,Versolato2011cjp,Dzuba2012pra,Portela2013hi,Gossel2013pra,Roberts2014pra}, for the high-precision atomic clock research~\cite{Dzuba2000pra,Sahoo2007pra,Sahoo2009pra,Versolato2011pra,Holliman2019pra} and quantum information processing~\cite{Fan2019prl}. Up to now, $35$ Ra isotopes have been discovered~\cite{Dammalapati16a}. Accurate knowledge of hyperfine-structure information of Ra$^{+}$ is the first step toward these high-precision researches.

Nuclear magnetic dipole moments $\mu_{I}$ of many Ra isotopes are known well. They have been directly measured for $^{213}$Ra and $^{225}$Ra at ISOLDE by the observation of Larmor precession of optically pumped atoms in a fast beam, and the $\mu_{I}$ of other odd isotopes were deduced from that~\cite{Arnold}. However, there is no direct measurement of the electric quadrupole moments $Q$ of Ra isotopes to date. The widely referenced values for the electric quadrupole moments $Q$ of Ra$^{+}$ were based on a semi-empirical analysis of HFS energy of the transition $7s_{1/2}\rightarrow7p_{3/2}$~\cite{Neu1988epd}. Especially noteworthy is that the semi-empirical $Q$ value~\cite{Neu1988epd} results in a about 40\% difference between the experimental and theoretical values of the electric quadrupole HFS constant of the $6d_{3/2}$ state of $^{211}$Ra$^{+}$~\cite{Sahoo2007pra,Versolato2011pla}. Although the magnetic dipole and electric quadrupole HFS constants of the first low-lying states of Ra$^{+}$ were measured and calculated, the magnetic octupole HFS constants and nuclear magnetic octupole moment $\Omega$ of Ra$^{+}$ have not been explored yet.  With the advent of modern spectroscopic technologies, the hyperfine splitting of $^{133}$Cs~\cite{Gerginov2003prl}, $^{137}$Ba$^{+}$~\cite{Lewty2013pra}, and $^{171}$Yb~\cite{Singh2013pra} were measured precisely, from which the contributions from magnetic octupole interactions were extracted, and the nuclear octupole moment $\Omega$ were determined by combining with the theoretical HFS constant $C$. Thus, it is also very interesting to perform a systematic investigation of magnetic octupole HFS constants of Ra$^{+}$ isotopes.

In this work, we employed the relativistic coupled-cluster method within single- and double-excitation approximation(CCSD), based on an even-tempered Gaussian-type basis set, to calculate the HFS constants of Ra$^+$ isotopes due to the interactions of nuclear magnetic dipole, electric quadrupole, and magnetic octupole moments and the electron. To understand the role of electron correlation in the calculations, the intermediate results from lower-order many-body perturbation-theory (MBPT) and the linear version of coupled-cluster with single and double approximation (LCCSD) are also presented. The electrical quadrupole moment $Q$ of $^{209,211,221,223}$Ra were extracted out by combining our results with the measured HFS constants $B$ of the $7p_{3/2}$ or $6d_{3/2}$ states. Using the HFS constants $C$ of Ra$^+$ and the nuclear shell model, the effect of magnetic octupole moment $\Omega$  on the hyperfine structure has also been evaluated. The following section presents a brief outline of the hyperfine structure theory and the coupled-cluster method for one-electron attachment processes. Numerical results presented in section III are compared with available theoretical and experimental data. Finally, a summary is given in section IV.

\section{THEORETICAL METHOD}
\subsection{The hyperfine structure theory}
The hyperfine interaction Hamiltonian for a relativistic electron takes the form~\cite{Schwartz1955PR}
\begin{small}\begin{equation}
H_{\mathrm{HFI}}=\sum_{k}T_{n}^{(k)}\cdot T_{e}^{(k)},
\end{equation}\end{small}
where $T_{n}^{(k)}$ and $T_{e}^{(k)}$ are the spherical tensor operators with rank $k$ ($k>0$) in the nuclear and electronic coordinates, respectively. The hyperfine states $|(\gamma IJ)F M_{F}\rangle$ with total angular momentum $\textbf{F}$=$\textbf{I}$+$\textbf{J}$ are represented by coupling an atomic angular momentum eigenstate $|\gamma J,M_{J}\rangle$ with a nuclear angular momentum eigenstate $|I,M_{I}\rangle$, where $\gamma$ encapsulating the remaining electronic quantum numbers.
Restricted to $k\leq3$, the first-order correction $W_{F, J}^{(1)}$ of hyperfine interaction to the energy can be parameterized in terms of the HFS constants $A$, $B$, $C$, and angular momentum coefficients
\begin{small}
\begin{equation}\begin{aligned}
W_{F, J}^{(1)}=&\sum_{k_{1}=1}^{3}\langle (\gamma IJ) F M_{F}|T_{n}^{(k_{1})}\cdot T_{e}^{(k_{1})} |(\gamma IJ) F M_{F}\rangle\\
=&\underbrace{\frac{1}{2}KA}_{M1:\,{k_{1}=1}}+\underbrace{\frac{1}{2}\frac{3K(K+1)-4I(I+1)J(J+1)}{2I(2I-1)2J(2J-1)}B}_{E2:\,{k_{1}=2}}\\
&+\frac{1}{[I(I-1)(2 I-1) J(J-1)(2 J-1)]}\times\left\{(5/4){K}^{3}\right.\\
&+5{K}^{2}+{K}\times[-3 I(I+1)
\times J(J+1)+I(I+1)
\\&\underbrace{+J(J+1)+3]-5 I(I+1) J(J+1)\}C\qquad\qquad}_{M3:\,{k_{1}=3}}.
\end{aligned}\end{equation}
\end{small}
Here the constants $A$, $B$, and $C$ are defined as
\begin{small}
\begin{equation}\begin{aligned}
A &=\mu_{N}\frac{\mu_{I}}{I}\frac{\langle {\gamma J|}\big |{ T_{e}^{(1)}}\big |{|\gamma J}\rangle}{\sqrt{J(J+1)(2J+1)}}\\
\end{aligned}\end{equation}
\end{small}
\begin{small}
\begin{equation}\begin{aligned}
B &=2Q \left [\frac{2J(2J-1)}{(2J+1)(2J+2)(2J+3)}\right ]^{1/2}{\langle {\gamma J|}\big |{ T_{e}^{(2)}}\big |{|\gamma J}\rangle}\\
&=Q R^{E2}(\gamma J),
\end{aligned}\end{equation}
\end{small}
and
\begin{small}
\begin{equation}\begin{aligned}
C &=\Omega_{I} \left [\frac{J(2J-1)(J-1)}{(J+1)(J+2)(2J+1)(2J+3)}\right ]^{1/2}{\langle {\gamma J|}\big |{ T_{e}^{(3)}}\big |{|\gamma J}\rangle}\\
&=\Omega_{I} R^{M3}(\gamma J),
\end{aligned}\end{equation}
\end{small}
where $K=F(F+1)-I(I+1)-J(J+1)$, and $\mu_{N}$ is the nuclear Bohr magneton. We have defined two ratios, \textit{i.e.}$R^{E2}(\gamma J)$=$B/Q$ and $R^{M3}(J)$=$C/\Omega_{I}$. They can be determined directly from theoretical calculations.

The contributions from the magnetic dipole$-$magnetic dipole ($M1-M1$) and magnetic dipole$-$electric quadrupole ($M1-E2$) second-order hyperfine interactions are generally on the same order as the magnetic octupole contribution, thus we also take into account these two second-order contribution terms. The second-order correction $W_{F, J}^{(2)}$ can be written as~\cite{Beloy2008PRA}
\begin{small}
\begin{equation}\label{eq:22}\begin{aligned}
W_{F, J}^{(2)}=&\sum_{\gamma J^{\prime}}\sum_{k_{1},k_{2}} \frac{\langle(\gamma I J^{\prime}) F M_{F}|T_{n}^{(k_{1})}\cdot T_{e}^{(k_{1})}|(\gamma IJ)F M_{F}\rangle}{E_{\gamma J}-E_{\gamma^{\prime} J^{\prime}}}\\&\times\langle(\gamma I J^{\prime}) F M_{F}|T_{n}^{(k_{2})}\cdot T_{e}^{(k_{2})}|(\gamma IJ)F M_{F}\rangle\\
\approx&\underbrace{\sum_{J^{\prime}}\left|\left\{\begin{array}{lll}
F & J & I \\
k_{1} & I & J^{\prime}
\end{array}\right\}\right|^{2} \eta}_{M1-M1:\,k_{1}=k_{2}=1}
+\underbrace{\sum_{J^{\prime}}\left\{\begin{array}{lll}
F & J & I \\
k_{1} & I & J^{\prime}
\end{array}\right\}\left\{\begin{array}{lll}
F & J & I \\
k_{2} & I & J^{\prime}
\end{array}\right\} \zeta}_{M1-E2:\,k_{1}=1,k_{2}=2},
\end{aligned}\end{equation}
\end{small}
where
\begin{small}
\begin{equation}
\eta=\frac{(I+1)(2I+1)}{I}{\mu^{2}_{I}}\frac{|{\langle {\gamma J^{\prime}|}\big |{ T_{e}^{(1)}}\big |{|\gamma J}\rangle}|^{2}}{E_{\gamma J}-E_{\gamma J^{\prime}}},
\end{equation}
\end{small}
\begin{small}
\begin{equation}\begin{aligned}
\zeta=&\frac{(I+1)(2I+1)}{I}  \sqrt{\frac{(2I+3)}{2I-1}}\\
      &\times{\mu_{I}}Q \frac{{\langle {\gamma J^{\prime}|}\big |{ T_{e}^{(1)}}\big |{|\gamma J}\rangle}{\langle {\gamma J^{\prime}|}\big |{ T_{e}^{(2)}}\big |{|J}\rangle}}{E_{\gamma J}-E_{\gamma J^{\prime}}}.
\end{aligned}\end{equation}
\end{small}

The single particle reduced matrix elements of the operators $T^{(1)}_e$, $T^{(2)}_e$, and $T^{(3)}_e$ are given by
\begin{eqnarray}\label{eq:88}
\langle{\kappa_{i}}&\|T^{(1)}_e\|&\kappa_{j}\rangle=-\langle-\kappa_{i}\|C^{(1)}\|\kappa_{j}\rangle
(\kappa_{i}+\kappa_{j})\notag\\
&\times&\int_{0}^{\infty}{dr\frac{P_{i}(r)Q_{j}(r)+P_{j}(r)Q_{i}(r)}{r^2}\times{F(r)}},
\end{eqnarray}
\begin{eqnarray}
\langle{\kappa_{i}}\|T^{(2)}_e\|\kappa_{j}\rangle&=&-\langle\kappa_{i}\|C^{(2)}\|\kappa_{j}\rangle\notag\\
&\times&\int_{0}^{\infty}{dr\frac{P_{i}(r)P_{j}(r)+Q_{j}(r)Q_{i}(r)}{r^3}},
\end{eqnarray}
and
\begin{eqnarray}
\langle{\kappa_{i}}\|T^{(3)}_e\|\kappa_{j}\rangle&=&-\frac{1}{3}\langle-\kappa_{i}\|C^{(3)}\|\kappa_{j}\rangle
(\kappa_{i}+\kappa_{j})\notag\\
&\times&\int_{0}^{\infty}{dr\frac{P_{i}(r)Q_{j}(r)+P_{j}(r)Q_{i}(r)}{r^4}},
\end{eqnarray}
where the relativistic angular-momentum quantum number $\kappa=\ell(\ell+1)-j(j+1)-1/4$, $P$ and $Q$ are the large and small radial components of Dirac wavefunction, respectively.

The $F(r)$ in Eq.(\ref{eq:88}) is a factor describing the correction of the finite nuclear magnetization distribution called Bohr-Weisskopf effect (BW)~\cite{Bohr1950pr}. When the point magnetization distribution (point) is used,  $F(r)=1$. It means that the BW effect is not considered in calculation. For the magnetic dipole hyperfine structure constants of heavy atomic system, the contribution of the BW correction cannot be ignored. The BW effect can usually be evaluated by approximating the magnetization distribution of the finite nucleus to a uniform sphere distribution model (sph) or a single-particle model (SP)~\cite{Roberts2020prl}. For the uniformly magnetized sphere (sph) model,
$F(r)$ can be written as
\begin{eqnarray}
  F_{sphere}(r)=\begin{cases}
  (\frac{r}{r_{N}})^{3}& r<r_{N}\\
  1& r\geq{r_{N}}
  \end{cases}.
  \end{eqnarray}
For the single-particle (SP) model, $F(r)$ is
\begin{eqnarray}
F(r)=F_{sphere}(r)\times\begin{cases}
1-\delta{F_{I}\ln(\frac{r}{r_{N}})}& r<r_{N}\\
1& r\geq{r_{N}}
\end{cases},
\end{eqnarray}
with
\begin{eqnarray}
\delta{F_{I}}=\begin{cases}
\frac{3(2I-1)}{8(I+1)}\frac{4(I+1)g_{L}-g_{S}}{g_{I}I}
&I=L+1/2\\
\frac{3(2I+3)}{8(I+1)}\frac{4Ig_{L}+g_{S}}{g_{I}I}&I=L-1/2\\
\end{cases},
\end{eqnarray}
where $r_{N}=\sqrt{5/3}r_{\rm rms}$ ($r_{\rm rms}$ is the root-mean-square radius of the nucleus. ).  $I$, $L$, and $S$ are the total, orbital, and spin angular momentum of the unpaired nucleon. Taking $g_{L}=1$ for the proton and $g_{L}=0$ for the neutron, the effective spin g factor, $g_{S}$, is determined from the experimental value $g_{I}=\mu/(\mu_{N}I)$ using the formula
\begin{equation}
g_{I}=\frac{1}{2}[g_{L}+g_{s}+(g_{L}-g_{s})\frac{L(L+1)-S(S+1)}{I(I+1)}].
\end{equation}
According to the Ref.~\cite{Shabaev1995pra,Shabaev1997pra}, the SP model can make a rough assessment of BW correction, but it can also be further improved by including the spin-orbit interaction and founding the nucleon wavefunction in a Woods-Saxon potential (SP$-$WS). Present work pays more attention to the high-order hyperfine interaction, ie. $B/Q$ and $C/\Omega$, so here we only use the sphere model and SP model to discuss the BW correction of the HFS constant $A$. The required parameters of radium isotopes can be found in Ref.~\cite{Angeli2004adndt}.
\subsection{A brief description of the relativistic coupled-cluster approach}
In the coupled-cluster theory, the atomic wave function $|\Psi_{\upsilon}\rangle$ for a single valence($\upsilon$) open-shell system is given by
\begin{small}
\begin{equation}
|\Psi_{\upsilon}\rangle= { e^{S} }|\Phi_{\upsilon}\rangle.
\end{equation}
\end{small}
$|\Phi_{v}\rangle$ is the reference state which is set as the zero-order Dirac-Fock wavefunction in present work, and $S$ represents the cluster operator. Within the single- and double-excitation approximation, the cluster operator $S$ can be partitioned into
\begin{eqnarray}
S&=&S^{(0,0)}+S^{(0,1)}\notag\\
 &=&S^{(0,0)}_1+S^{(0,0)}_2+S^{(0,1)}_1+S^{(0,1)}_2,
\end{eqnarray}
where $S^{(0,0)}$ and $S^{(0,1)}$ correspond to the excitation from core and from valence electrons, respectively. The expected value of one-particle operator $O$ for the state $|\Psi_{\upsilon}\rangle$ can be evaluated using the following expression
\begin{small}
	\begin{equation}
{\bar{O}_{\upsilon}}=\frac{\langle{\Psi_{\upsilon}}|{O}|{\Psi_{\upsilon}}\rangle}{\langle{\Psi_{\upsilon}}|{\Psi_{\upsilon}}\rangle}=\frac{\langle{\Phi_{\upsilon}}|{e^{S^{\dagger}}Oe^{S}}|{\Phi_{\upsilon}}\rangle}{\langle{\Phi_{\upsilon}}|{e^{S^{\dagger}}e^{S}}|{\Phi_{\upsilon}}\rangle}.
	\end{equation}
\end{small}
where the numerator and the denominator $e^{S^{\dagger}}Oe^{S}$ are not terminating. In previous works~\cite{Tang2019jpb,Lou2019aps}, only linear terms of cluster operators were included in property calculations,
 {\it i.e.},
\begin{eqnarray}
\label{lcc1}
e^{S\dag}O{e^{S}}\approx&O&+\{{O}{S^{(0,0)}_{1}}+{\rm c.c.}\}+\{{O}{S^{(0,1)}_1}+{\rm c.c.}\}\notag\\
                 &+&\{{O}{S^{(0,1)}_{2}}+{\rm c.c.}\}+\{S^{(0,0)\dag}_1O{S^{(0,1)}_1}+{\rm c.c.}\}\notag\\
                 &+&S^{(0,0)\dag}_1O{S^{(0,0)}_1}+\{S^{(0,0)\dag}_1O{S^{(0,0)}_2}+{\rm c.c.}\}\notag\\
                 &+&\{S^{(0,0)\dag}_1O{S^{(0,1)}_2}+{\rm c.c.}\}+S^{(0,0)\dag}_2O{S^{(0,0)}_2}\notag\\
                 &+&\{S^{(0,0)\dag}_2O{S^{(0,1)}_2}+{\rm c.c.}\}+S^{(0,1)\dag}_1O{S^{(0,1)}_1}\notag\\
                 &+&\{S^{(0,1)\dag}_1O{S^{(0,1)}_2}+{\rm c.c.}\}+S^{(0,1)\dag}_2O{S^{(0,1)}_2}\notag\\
\end{eqnarray}
and
\begin{eqnarray}
\label{lcc2}
e^{S\dag}{e^{S}}\approx &1&+{S^{(0,0)\dag}_1}{S^{(0,0)}_1}+{S^{(0,1)\dag}_1}{S^{(0,1)}_1}\notag\\
                        &+&{S^{(0,0)\dag}_2}{S^{(0,0)}_2}+{S^{(0,1)\dag}_2}{S^{(0,1)}_2}
\end{eqnarray}
where ${\rm c.c.}$ stands for the complex conjugate part. For the LCCSD method, Eq.~\ref{lcc1}
and Eq.~\ref{lcc2} are exact.  But this approximation ignores the so-called ``dressing" from the nonlinear terms for the CCSD method.
To account for the contribution from the main non-linear terms of cluster operators, we use an iteration strategy as proposed in Ref.~\cite{Mani2010pra}. The basic idea of this strategy is that the $e^{S^{(0,0)\dagger}}Oe^{S^{(0,0)}}$ was expanded in terms of effective one-body, two-body, and three-body terms. These terms are obtained by a self-consistently iteration calculation and stored as an intermediate block. Then these intermediate blocks are constructed with $S^{(0,1)\dagger}$ and $S^{(0,1)}$ to the final diagrams. It is worth noting that one should avoid the repetition of any diagram in the iterative procedure. In this work, we calculate the expected value using three coupled-cluster approximations: $\rm LCCSD$, $\rm CCSD_{L}$, and $\rm CCSD$.  $\rm LCCSD$ and $\rm CCSD_{L}$ represent that the expected values are computed using Eq.~\ref{lcc1}
and Eq.~\ref{lcc2} with the cluster amplitudes from LCCSD and CCSD method respectively. $\rm CCSD$ corresponds to the self-consistently iteration calculation.

 \subsection{Computational outline}

The zero-order wave-function $|\Phi_{v}\rangle$ is obtained by the Dirac-Fock calculation. The no-pair Dirac Hamiltonian is set as the starting point. Breit interaction is considered at the same foot as coulomb interaction.  Different from the previous works~\cite{Tang2017pra,Tang2019jpb}, the lowest-order quantum electrodynamics (QED) correction is also included in Dirac-Fock formalism using a simple radiative potential proposed by Flambaum and Ginges~\cite{Flambaum2005pra}. The radiative potential can be
written as
\begin{equation}
\label{qed1}
\Phi_{\rm rad}(r)=\Phi_{U}(r)+\Phi_{g}(r)+\Phi_{f}(r)+\Phi_{l}(r)+\frac{2}{3}\Phi^{\rm Simple}_{WC}(r)
\end{equation}
where $\Phi_{U}(r)$ is the Uehling potential, $\Phi_{g}(r)$ is the magnetic form-factor contribution, $\Phi_{f}(r)$ is the high-frequency electric form-factor contribution, $\Phi_{l}(l)$ is the low-frequency electric form-factor contribution, and
$\Phi^{\rm Simple}_{WC}(r)$ is the Wichmann-Kroll potential, $\Phi_{U}(r)$ and $\Phi^{\rm Simple}_{WC}(r)$ are for the vacuum polarization, and the other three parts correspond to self-energy corrections. The expression of these effective operators can be found in Ref.~\cite{Ginges2016pra,Ginges2016jpb}. To check our QED code, we reproduced almost all results in Ref.~\cite{Ginges2016pra,Ginges2016jpb}.

In this work, we employed a finite basis set composed of even-tempered Gaussian-type functions to expand the Dirac radial wave function, as in Ref.\cite{Chaudhuri1999pra}. To avoid the spurious state and variational collapse problem, the kinetic balance condition between the large and the small components is used. The Gaussian-type function has the form
\begin{small}
 \begin{equation}
G_{i,k}=\mathcal N_{i}r^{n_{k}}e^{-\eta_{i}r^{2}},
\end{equation}
\end{small}
where $\mathcal N_{i}$ is the normalization factor, $n_{k}=l-1$, and $\eta_{i}=\eta_{0}\xi^{i-1}$. Then the large and small components of the orbital can be expressed as
\begin{small}
 \begin{equation}
\begin{array}{l}
P(r)=\sum \limits_{i=1}^{N}C_{i}^{P_{\kappa}}G_{i,\kappa}^{r}, \\
Q(r)=\sum \limits_{i=1}^{N}C_{i}^{Q_{\kappa}}\mathcal N_{i}^{Q}( \frac{d}{dr}+ \frac{\kappa}{r}  )G_{i,\kappa}^{r},
\end{array}
\end{equation}
\end{small}
with
\begin{small}
\begin{equation}
\mathcal N_{i}^{Q}=\sqrt{\frac{\alpha_{i}}{2n_{\kappa}-1}[4(\kappa^{2}+\kappa-n_{\kappa})-1]}.
\end{equation}
\end{small}

The Fermi nuclear distribution was employed to describe the Coulomb potential between electrons and the nucleus. All the core orbitals and virtual orbitals with energies smaller than 10000 a.u. are included in correlation calculations. Table~\ref{basis} lists the Gauss basis parameters. $N$ is the number of basis set for each symmetry, and $N_{c}$ and $N_{v}$ represent the number of core orbitals and virtual orbitals respectively.
\begin{table}[H]
\newcommand{\RNum}[1]{\uppercase\expandafter{\romannumeral #1\relax}}
\caption{The parameters of the Gauss basis set. $N$ is the number of basis set for each symmetry. $N_{c}$ and $N_{v}$ represent
the number of core orbitals and virtual orbitals, respectively. }\label{basis}
\renewcommand\tabcolsep{2.0pt}
\begin{ruledtabular}
\begin {tabular}{lcccccccc}
  &$s$&$p$&$d$&$f$&$g$&$h$&$i$&$j$
  \\   \hline
  $\eta_{0}\times10^{3}$ &0.42&0.42&0.42&0.57&7.6&86&95&96\\
  $\xi $ &1.71   &1.69   &1.71   &1.74   &2.0   &2.0  &2.0   &2.0\\
  $N$      &55     &45     &42     &35     &25    &20 &15    &10  \\
  $N_{c}$  &6      &5      &3      &1      &0     &0 &0     &0  \\
  $N_{v}$  &29     &28     &27     &25     &17    &14 &13    &10  \\
\end{tabular}
\end{ruledtabular}
\end{table}

\section{RESULTS AND DISCUSSION}

\subsection{Energies}
\begin{table*}[ht]\small
\begin{threeparttable}
\newcommand{\RNum}[1]{\uppercase\expandafter{\romannumeral #1\relax}}
\caption{ Energy levels of Ra$^{+}$ in ${\rm cm^{-1}}$. ${\rm E_{DF}}$ denotes the lowest-order Dirac-Fock energy.
 ${\rm E_{MBPT(2)}}$, ${\rm E_{LCCSD}}$, ${\rm E_{CCSD}}$, and ${\rm E_{CCSD+QED}}$ are the energies obtained using second-order MBPT, LCCSD, CCSD, and CCSD results with QED correction approximations, respectively. ${\rm \delta_{i}(\%)}$, where i=0, 1, 2, 3, 4 shows the differences between corresponding  calculation results and experimental values.  }\label{energy1}
\begin{ruledtabular}
\begin {tabular}{lccccccccccc}
  Level&${\rm E_{DF}}$&${\rm E_{MBPT(2)}}$& ${\rm E_{LCCSD}}$ & ${\rm E_{CCSD}}$ &${\rm E_{CCSD+QED}}$ &${\rm E_{Expt.}}$\cite{NIST}&$\delta_{\rm 0}(\%)$&$\delta_{1}(\%)$ &$\delta_{\rm 2}(\%)$ & $\delta_{\rm 3}(\%)$& $\delta_{\rm 4}(\%)$
  \\   \hline
  $7s_{1/2}$&	$-$75872.96 &	$-$83359.54 &	$-$82506.42 &	$-$82077.91 &	$-$81989.19 &	$-$81842.49 &	$-$7.29 &	1.85 &	0.81 &	0.29 &	0.18\\
  $7p_{1/2}$&	$-$56825.75 &	$-$60979.06 &	$-$60821.14 &	$-$60518.69 &	$-$60521.04&	$-$60491.17 &	$-$6.06 &	0.81 &	0.55 &	0.05 &	0.05\\
 $ 7p_{3/2}$&	$-$52887.58 &	$-$56002.27 &	$-$55889.94 &	$-$55652.01 &	$-$55650.44&	$-$55633.64 &	$-$4.94 &	0.66 &	0.46 &	0.03  &	0.03   \\
  $6d_{3/2}$&	$-$62408.99 &	$-$71017.35 &	$-$70327.07 &	$-$69744.51 &	$-$69786.20&	$-$69758.22 &	$-$10.54&	1.80 &	0.82 &	$-$0.02 	 &	0.04  \\
  $6d_{5/2}$&	$-$61662.25 &	$-$69101.11 &	$-$68593.87 &	$-$68071.05 &	$-$68103.28&	$-$68099.50 &	$-$9.45 &	1.47 &	0.73 &	$-$0.04 	 &	0.00     \\		
\end{tabular}
\end{ruledtabular}
\end{threeparttable}
\end{table*}
\begin{table*}[ht]\small
  \begin{threeparttable}
  \newcommand{\RNum}[1]{\uppercase\expandafter{\romannumeral #1\relax}}
  \caption{ Comparison of our calculated energies ${\rm E_{CCSD+QED}}$ with other available theoretical and experimental data in ${\rm cm^{-1}}$.}~\label{energy2}
  \begin{ruledtabular}
  \begin {tabular}{lccccccccccc}
    Level&${\rm E_{CCSD+QED}}$&${\rm E_{CCSD}}$~\cite{Eliav1996pra}& ${\rm E_{SDpT}}$~\cite{Pal2009pra,Safronova2007pra} & ${\rm E_{CPM}}$~\cite{Dzuba2001pra} &${\rm E_{CPM+LD}}$~\cite{Dzuba2013pra}&${\rm E_{Expt.}}$~\cite{NIST}
    \\   \hline
     7s$_{1/2}$&$-$81989.19   &$-$82025 & $-$81508	& $-$81960& $-$81714& $-$81842.5 \\
     6d$_{3/2}$&$-$69786.20   &$-$69596 & $-$69488	& $-$70149& $-$69788& $-$69758.22\\
     6d$_{5/2}$&$-$68103.28   &$-$67936 & $-$67947	& $-$68449& $-$68045& $-$68099.51\\
     7p$_{1/2}$&$-$60521.04	&$-$60462 & $-$60326	& $-$60681& $-$60511& $-$60491.17\\
     7p$_{3/2}$&$-$55650.44	&$-$55629 & $-$55519	& $-$55734& $-$55625& $-$55633.64\\		
  \end{tabular}
  \end{ruledtabular}
  \end{threeparttable}
  \end{table*}

We have calculated the energies of some important low-lying states of Ra$^{+}$ using different models including Dirac-Fock(DF), second-order many-body perturbation theory (MBPT(2)), LCCSD, and CCSD calculations, and the predicted energies are labeled by ${\rm E_{DF}}$, ${\rm E_{MBPT(2)}}$, ${\rm E_{LCCSD}}$, and ${\rm E_{CCSD}}$, respectively. We also evaluated the correction of the QED effect to the calculation of CCSD values labeled by ${\rm E_{CCSD+QED}}$. The results of the first five states are listed in table~\ref{energy1} and compared with experimental values from NIST~\cite{NIST} labeled as ${\rm E_{Expt.}}$. From table~\ref{energy1}, one can easily find that: (i) There are obvious differences between DF and CCSD energies indicating significant contributions from electron correlation effects that are not included in DF calculations. The largest contribution is approximately $10\%$ occurring at $6d$ states. (ii) The differences between CCSD results and the experimental values are no more than $0.1\%$ except for the ground state, showing a much better agreement than MBPT(2) and LCCSD results for all states. This phenomenon indicates that the inclusion of nonlinear terms of the cluster operators are of crucial importance in obtaining very accurate energy levels. (iii) It is necessary to consider the QED effect for the energy improvement of the low energy states, especially for the ground state, the energy improvement is $0.1\%$.

There are also some other theoretical calculations about radium ion energies~\cite{Eliav1996pra, Safronova2007pra,Pal2009pra,Dzuba2001pra,Dzuba2013pra}. Table~\ref{energy2} presents a comprehensive comparison of our calculated energies with other $ab$ $initio$ results. Elaiv \textit{et al.}~\cite{Eliav1996pra} calculated
the energies of the first low-lying states of Ra$^{+}$ and Ra using relativistic CCSD based on DCB Hamiltonian. Our method is the same as theirs, but we use a larger basis set and partial waves.
 Safronova \textit{et al.}~\cite{Safronova2007pra} and Pal~\textit{et al.}~\cite{Pal2009pra} carried out their calculations using the all-order many-body perturbation theory, in which the linear single- and double-excitation terms are included and a part of triple-excitation are also considered using third-order MBPT corrections (SDpT). The method adopted by Dzuba \textit{et al.}~\cite{Dzuba2001pra} is called the correlation potential method(CPM). In Ref.~\cite{Dzuba2013pra}, the ladder diagrams were added to the correlation potential method (CPM+LD), which significantly improved the level of accuracy for the lowest $D$ states. From Table~\ref{energy2}, one can notice that our total CCSD results, about 0.2\% better than the SDpT results~\cite{Pal2009pra} with respect to the experimental values, are very close to the results by the CPM~\cite{Dzuba2001pra} and CPM+LD~\cite{Dzuba2013pra} which also include the QED corrections.

\subsection{Hyperfine structure constants A of Ra$^{+}$ isotopes.}

\begin{table*}[]\small
  \begin{threeparttable}
  \newcommand{\RNum}[1]{\uppercase\expandafter{\romannumeral #1\relax}}
  \caption{The HFS constants $A$ (MHz) of $^{223}$Ra$^{+}$ at different correlation levels are given. The total results, \textit{ie.} CCSD+QED+(BW$-$SP)$_{\rm CC}$, are our recommended values. The contributions of higher partial waves, QED correction, BW correction and high-order correlation beyond CCSD are estimated respectively and summarized in brackets as the uncertainties of the recommended values. Some other $ab$ $initio$ theoretical and experimental results are also listed for comparison. }\label{A1}
  \begin{ruledtabular}
  \begin {tabular}{lcccccc}
  Method         &$7s_{1/2}$&$6d_{3/2}$&$6d_{5/2}$&$7p_{1/2}$&$7p_{3/2}$\\
  \hline
  DF            &2687   &52.9  &   19.3 	&438  &33.8 \\
  MBPT(3)       &3690   &80.6  &$-$19.3   &665  &53.5 \\
  LCCSD         &3663   &82.7  &$-$25.5   &689  &56.6 \\
  $\rm CCSD_{L}$&3621   &79.7  &$-$22.0   &666  &56.3 \\
  CCSD          &3608   &82.0  &$-$24.2   &666  &56.8 \\
  \multicolumn{6}{c}{Estimated BW and QED}\\
  (BW$-$sph)$_{\rm DF}$  &$-$66.9  &0.0      &0.0      &$-$1.62 &0.0     \\
  (BW$-$sph)$_{\rm CC}$  &$-$90.1  &$-$1.27&1.24   &$-$2.24 &$-$0.17      \\
  (BW$-$SP)$_{\rm DF}$   &$-$91.8  &0.0      &0.0      &$-$2.26 &0.0 \\
  (BW$-$SP)$_{\rm CC}$   &$-$123.5 &$-$1.73&1.70   &$-$3.12 &$-$0.24      \\
  QED                  &$-$59.7  &$-$0.54&0.68    &$-$2.72 &$-$0.03      \\
  \multicolumn{6}{c}{Estimated uncertainties}\\
  HO            & 45   &1.4      &2.1    &11.2      &1.2\\
  Basis         & 1.8  &0.2      &0.1    &0.4       &0.1 \\
  $\rm BW_c$    & 62   &$-$0.87  &0.85   &$-$1.56   &$-$0.12 \\
  $\rm QED_c$   & 30   &$-$0.27  &0.34   &$-$1.36   &$-$0.02 \\
  \multicolumn{6}{c}{Recommended values}\\
  Total      &3426(82)&79.7(1.7)&$-$21.8(2.3)&660(11.4)&56.5(1.2)\\
  SDpT~\cite{Pal2009pra}       &3450   &79.6 &$-$24.1 &671.5 &54.4	       \\
  CCSD(T)~\cite{Sahoo2007pra}  &3567.26&77.1 &$-$23.9 &      &	           \\
  Z-vector~\cite{Sasmal2017pra}&3446.3 &      &   	    &      &             \\
  Expt.1987~\cite{Wendt1987}  &       &      &         &667.1(2.1)&56.5(8)  \\
  Expt.1988~\cite{Neu1988epd}&3404.0(1.9) &  &         &          &         \\
  \end{tabular}
  \end{ruledtabular}
  \end{threeparttable}
  \end{table*}

Different isotopes of Ra have different hyperfine splittings resulted from their different nuclear spin.
$^{223}$Ra$^{+}$ as
an example, the HFS constant $A$ of the first five states from various methods are list in table~\ref{A1} in MHz.
 The theoretical results in the first five rows are obtained within the point magnetization distribution model. The following rows give the BW contributions from the sphere model and SP model at DF and CCSD levels, together with the QED correction from the radiation potential method. The following ``Estimated uncertainties" part provides the uncertainties due to the incomplete basis set (``Basis"), high-order correlation effect (``HO") beyond the CCSD method, BW correction(``$\rm BW_c$") and QED correction(``$\rm QED_c$").  Finally, the ``Total" row lists our recommended value obtained by CCSD+QED+(BW$-$SP)$_{\rm CC}$ with the uncertainty enclosed in parentheses. Some other $ab$ $initio$ theoretical and experimental results are also listed for comparison.

 In our calculation,  the ``Basis" uncertainty is estimated using the partial wave extrapolation strategy based on calculating all quantities under the conditions of $\ell_{max}= 4, 5, 6, 7$, respectively. The premise is that the number of Gaussian bases we use is large enough to satisfy our calculations. The contribution from high-order electron correlation effect beyond CCSD is difficult to effective estimating, but it is usually not greater than the contribution of the lower-order nonlinear terms (\textit{ie.}$\rm CCSD-LCCSD$) which makes up about 5\% of the total electron correlation. Therefore, we take the 5\% of the total electron correlation contribution, \textit{ie.}$\rm( CCSD-DF)\times5\%$ as the uncertainty. Ginges~\textit{et al.}~\cite{Ginges2017pra} investigated the BW correction for the HFS A of the ground state of some alkali-metal-like ions within sphere model, SP model, and SP$-$WS model respectively. The uncertainty of BW correction does not exceed 30\% of the maximum value of the two results from SP$-$WS model and the sphere model. For $^{223}$Ra$^{+}$, the BW correction obtained by SP model of the five states in table are greater than those from the sphere model, so we take 50\% of the total BW effect within the SP model as the conservatively estimated uncertainty. The radiation potential method here is reliable for estimating QED correction to binding energy, but is not rigorous for the short-distance operators such as hyperfine interaction operators. Compared with the more rigorous evaluation in Ref.~\cite{Ginges2017pra}, our calculation seems to overestimate the effect of QED and should be compensated with a large uncertainty, \textit{ie.} the QED$\times$50\%.

 From the table~\ref{A1}, we can find that the CCSD results show better agreement with experimental values than the LCCSD and the MBPT(3) method, and the MBPT(3) calculation yields the worst results. The inclusion of correlation effects by our CCSD method improved the results of $A$ from the DF values by $35\%$ in the $6d_{3/2}$ state and $180\%$ in the $6d_{5/2}$ state regardless of the sign, which suggests that the correlation effect plays a significant role in $d$ states. Moreover, the opposite sign of CCSD values and DF values of $6d_{5/2}$ states reveals a strong cancellation from different correlation effects. Compared with the experimental value of the ground state $7s_{1/2}$, 3404.0(1.9) MHz, our CCSD result, 3608 MHz, is overestimated by about 6\%. After further considering BW and QED corrections (both are negative values), the result changes to 3426(82) MHz, then the difference is significantly reduced to 0.6\%, showing that these two corrections are extremely important to further improve the accuracy of the A for the ground state, moreover, the contribution of BW corrections seems in an even more dominant position than the QED radiative correction. The BW effect also plays an important role in $d$ states. Especially for $6d_{5/2}$, the total BW correction obtained by the SP model accounts for 7\% of the CCSD result, which is 2\% larger than that obtained from the sphere model. Consequently, the assessment of the BW is necessary for the state with a relatively large correlation effect, and a more accurate nuclear model is also required in this process. Some interesting features about the BW correction of the states with high angular momentum can also be found in the table. For $7p_{3/2}$, $6d_{3/2}$, and $6d_{5/2}$ states, whether within the sphere distribution model or the SP model, the BW corrections to DF are zero, and the total BW corrections only come from the calculation of the correlation effect, which indicates that the accurate calculation of the correlation effect is very important for the investigation of the BW correction on these states. Overall, our calculations show that the BW and QED corrections are important for the HFS constant A of Ra$^{+}$. To further obtain a more accurate wave function of different radium isotope ion and reduce the current uncertainty, it is necessary and important to  optimize the magnetization distribution model and perform a rigorous calculation of the QED effect.

 The present CCSD, SDpT~\cite{Pal2009pra}, CCSD(T)~\cite{Sahoo2007pra}, Z-vector~\cite{Sasmal2017pra} methods in Table~\ref{A1} are all based on relativistic coupled-cluster-theory, though they have some different treatments for electronic correlation effects. In the SDpT method, other nuclear magnetic distribution models are used to consider the BW correction, while there is no description of the BW effect in the CCSD(T) and Z-Vector methods, and none of these three methods evaluate the contribution of QED effect. Compared with other results, our final value of the $7s_{1/2}$ and $7p_{3/2}$ states including the BW and QED corrections are the closest to the experimental value. The experimental value of the $7p_{1/2}$ state is between our total value and the SDpT value, which indicates that there may be some offset among the (BW + QED) corrections, non-linear terms and the higher-order excitations. We find that all the theoretical HFS constant A of ground state are larger than the experimental value.

\subsection{Evaluation of the electric quadrupole moment Q for different Ra$^+$ isotopes.}

Due to the absence of direct measurements of electric quadrupole moments $Q$, extracting the $Q$ value out by combining the measured HFS constant $B$ and calculated electric field gradient or $R^{E2}(\gamma J)$ is usually considered to be convenient and economical. From this point, a very precise measurement of the hyperfine splitting and accurate theoretical determination of the ratio value $R^{E2}(\gamma J)$ is needed to extract the $Q$ value accurately. The validity of our CCSD calculation has been verified by the above analysis for energies and HFS constants $A$. In the same theoretical framework, we calculated the ratio $R^{E2}(\gamma J)$ (in MHz/b). The $R^{E2}(\gamma J)$ of $7p_{3/2}$, $6d_{3/2}$ and $6d_{5/2}$ states from various
methods are listed in Table~\ref{B}. For this quantity, the uncertainty is from the incomplete basis set and the high-order correlation effect beyond CCSD method. The contributes of QED correction is almost zero. Our recommended result of 309(5) MHz/b for $R^{E2}$ of 6d$_{3/2}$ state is very close to the value of 306.12 MHz/b in Ref.~\cite{Sahoo2007pra}. It implies that the 40\% difference between the experimental and theoretical values of the HFS constant $B$ of the $6d_{3/2}$ state of $^{211}$Ra$^{+}$~\cite{Sahoo2007pra,Versolato2011pla} may be caused by inaccurate estimation of $Q$ and prompted us to evaluate $Q$ in combination with available experimental HFS constant $B$ and our CCSD $R^{E2}(\gamma J)$.
\begin{table}[H]\small
\begin{threeparttable}
\newcommand{\RNum}[1]{\uppercase\expandafter{\romannumeral #1\relax}}
\caption{ $R^{E2}(\gamma J)$ (in MHz/b) of $7p_{3/2}$, $6d_{3/2}$ and $6d_{5/2}$ states at different levels of correlation. Our recommended value is listed in the $\rm{``Final"}$ column, where the uncertainty is in parentheses.}\label{B}
\begin{ruledtabular}
\begin {tabular}{lcccccccccc}
\toprule
  Level&${\rm DF}$&${\rm LCCSD}$ & ${\rm CCSD_L}$&${\rm CCSD}$&Final\\
   \hline
$7p_{3/2}$	&	398 	&	714 &692  &693  &693(15)	\\
$6d_{3/2}$	&	213 	&	309 &311  &309  &309(5)	    \\
$6d_{5/2}$	&	227 	&	386 &386  &384  &384(8)	    \\			
\end{tabular}
\end{ruledtabular}
\end{threeparttable}
\end{table}
\begin{table*}[ht]
  \begin{threeparttable}
  \newcommand{\RNum}[1]{\uppercase\expandafter{\romannumeral #1\relax}}
  \caption{ The electric quadrupole moment $Q$ (in b) of $^{209,211,221,223}$Ra. $Q_{\rm ccsd}$ are our values. ${Q_{\rm semi}}$ are from Ref.~\cite{Neu1988epd}. Uncertainties are
  given in parentheses. The subscript $\rm{``E"}$ and $\rm{``T"}$ represent the uncertainty coming from the experiment value $ B_{\rm Expt.}$  and the theoretical calculation $R^{E2}(\gamma J)$, respectively.}\label{QQ}
  \begin{ruledtabular}
  \begin {tabular}{lcccc}
   \multicolumn{1}{c}{}
   &\multicolumn{2}{c}{6d\,$^{2}$D$_{3/2}$}
   &\multicolumn{2}{c}{7p\,$^{2}$P$^{o}_{3/2}$}\\
  \cline{2-3}\cline{4-5}
  \multicolumn{1}{c}{}
  &\multicolumn{1}{c}{$^{209}$Ra$^{+}$}
  &\multicolumn{1}{c}{$^{211}$Ra$^{+}$}
  &\multicolumn{1}{c}{$^{221}$Ra$^{+}$}
  &\multicolumn{1}{c}{$^{223}$Ra$^{+}$}\\
  \hline
  ${B_{\rm Expt.}}$ &104(38)   &103(6) &1364.2(5.1)&864.8(1.9)  \\
  ${Q_{\rm ccsd}}$  &0.337$(123)_{\rm{E}}$$(6)_{\rm{T}}$      &0.333$(19)_{\rm{E}}$$(6)_{\rm{T}}$ &1.968$(7)_{\rm{E}}$$(34)_{\rm{T}}$        &1.248$(3)_{\rm{E}}$$(22)_{\rm{T}}$    \\
  ${Q_{\rm semi}}$  &0.40$(2)_{\rm{E}}$$(4)_{\rm{T}}$          &0.48$(2)_{\rm{E}}$$(4)_{\rm{T}}$ &1.978$(7)_{\rm{E}}$$(106)_{\rm{T}}$	   &1.254$(3)_{\rm{E}}$$(66)_{\rm{T}}$  \\
  \end{tabular}
  \end{ruledtabular}
  \end{threeparttable}
  \end{table*}

Since the HFS constants $B$ are only measured for $7p_{3/2}$ state of $^{221,223}$Ra$^{+}$ and $6d_{3/2}$ state of $^{209,211}$Ra$^{+}$, we evaluated the electric quadrupole moment
$Q$ for these states, labeled as $ Q_{\rm ccsd}$, and compared with a semi-empirical analysis, $ Q_{\rm semi}$, from Ref.~\cite{Neu1988epd} in table~\ref{QQ}. The experimental values of electric quadrupole HFS constants, $B_{\rm Expt.}$, are also listed in the table. The equation used to evaluate the electric quadrupole moment $Q$ is
\begin{small}\begin{equation}
Q = \frac{B_{\rm Expt.}}{R^{E2}}(\gamma J).
\end{equation}\end{small}
Obviously, the accuracy of resulting $Q$ is dependent on the accuracy of $B_{\rm Expt.}$ and $R^{E2}$. We use the results in Table~\ref{QQ}. In Table~\ref{QQ}, uncertainties are given in parentheses. The subscript $\rm{``E"}$ and $\rm{``T"}$ represent the uncertainties from $B_{\rm Expt.}$ and $R^{E2}(\gamma J)$, respectively. Our $Q_{\rm ccsd}$ are consistent with $Q_{\rm semi}$ for $^{221,223}$Ra with a difference of only $0.5\%$, but are significantly smaller than $Q_{\rm semi}$ in the cases of $^{209,211}$Ra. One should also note that the experimental HFS constants $B$ for the $6d_{3/2}$ state of $^{209,211}$Ra$^+$ have larger uncertainties.

 Furthermore, according to the ratio values provided by present work, if the HFS constant $B$ of a state $|{\gamma^{\prime} J^{\prime}}\rangle$ of other Ra$^{+}$ isotopes, denoted by $B_{1}(\gamma^{\prime} J^{\prime})$, are measured, we can easily obtain their electric quadrupole moment $Q_{1}$ according to the following relations:
 \begin{small}\begin{equation}
\frac{B_{1}(\gamma^{\prime} J^{\prime})}{B_{2}(\gamma J)}=\frac{Q_{1} R^{E2}_{1}(\gamma^{\prime} J^{\prime})}{Q_{2} R^{E2}_{2}(\gamma J)}
 \end{equation}\end{small}
where the $B_{2}(\gamma J)$ is the experimental HFS constant $B$ for a given state $|\gamma J\rangle$ and $Q_{2}$ is the CCSD value for a certain Ra$^{+}$ isotope.

\subsection{Appraising nuclear octupole moment contributions to the hyperfine structures.}
In this section, we use the nuclear octupole moment $\Omega$ from the single-particle shell model to roughly evaluate the HFS constant $C$ of some low-lying states in Ra$^{+}$. We expect to obtain the typical orders of magnitudes of the hyperfine splitting caused by magnetic octupole hyperfine interaction.
This process is necessary to know the precisions required of these hyperfine spectral measurements to extract the nuclear octupole moment $\Omega$ of Ra. In the absence of reference values for $\Omega$ of Ra, we can estimate it crudely by the single-particle model:
\begin{equation}
\begin{aligned}
\Omega^{\mathrm{sp}}=& \mu_{N}\left\langle r^{2}\right\rangle \frac{3}{2} \frac{(2 I-1)}{(2 I+4)(2 I+2)} \\
& \times\left\{\begin{array}{c}
(I+2)[(I-\frac{3}{2}) g_{L}+g_{S}],\text { for } I=L+\frac{1}{2}\\
(I-1)[(I+\frac{5}{2}) g_{L}-g_{S}],\text { for } I=L-\frac{1}{2}
\end{array}\right.& \\
\end{aligned}
\end{equation}

For even-odd (even number of protons, odd number of neutrons) nuclei Ra, $g_{L}$ =0, $g_{S}$ =$-$3.826. Here we set $^{223}$Ra as an example, possessing nuclear ground states with a spin of $I=3/2$, and  an orbital momentum $L=2$. Then we may estimate the octupole moment to be
\begin{equation}
\Omega^{\mathrm{sp}}=0.164 \mu_{N}\left\langle r^{2}\right\rangle \approx 0.0525 \mu_{N} \times \mathrm{b}.
\end{equation}
The root-mean-square radius {$\left\langle r^{2}\right\rangle$}$^{1/2}$ of the nucleus of $^{223}$Ra is 5.6602 $fm$. The ratio of magnetic octupole hyperfine constants to magnetic octupole moments $R^{M3}(\gamma J)$ are listed in Table~\ref{C}.
\begin{table}[H]\small
  \begin{threeparttable}
  \newcommand{\RNum}[1]{\uppercase\expandafter{\romannumeral #1\relax}}
  \caption{ $R^{M3}(\gamma J)$ (in kHz/($\mu_{N}$$\times$b)) of $7p_{3/2}$, $6d_{3/2}$ and $6d_{5/2}$ states at different levels of correlation. Our recommended value is listed in the $\rm{``Final"}$ column, where the uncertainty is in parentheses.}\label{C}
  \begin{ruledtabular}
  \begin {tabular}{lcccccccccc}
  \toprule

    Level&${\rm DF}$&${\rm LCCSD}$ & ${\rm CCSD_L}$&${\rm CCSD}$&Final \\   \hline
  $7p_{3/2}$	&		26.2 	&	42.9 	&	41.4 	&41.4 &41.4(8)\\
  $6d_{3/2}$	&		6.71 	&	9.73 	&	9.64 	&9.66&9.66(15)\\
  $6d_{5/2}$	&		1.82 	&$-$6.17 	&$-$5.99 	&$-$6.23&$-$6.23(40)\\			
  \end{tabular}
  \end{ruledtabular}
  \end{threeparttable}
  \end{table}
It is also seen in the table that the trends in the correlation effects for the calculations of $A$ and $C$ are almost the same ignoring the $d_{5/2}$ states which have a larger uncertainty. Comparing the magnitudes of the calculated $C/\Omega$ values among the given states, the $C^{7p_{3/2}}$ is large:

\begin{small}
\begin{equation}
C^{7p_{3/2}}=2172\ \mathrm{Hz}.
\end{equation}
\end{small}

For the $7p_{3/2}$ state, the hyperfine structure intervals $\delta W_{F}=W_{F,7p_{3/2}}-W_{F-1,7p_{3/2}}$ can be expressed in terms of these constants as

\begin{small}
\begin{equation}
\label{eq1}\delta W_{1}=A^{7p_{3/2}}-B^{7p_{3/2}}+56C^{7p_{3/2}}+\frac{\eta^{7p_{3/2}}}{36}-\frac{\sqrt{5}\zeta^{7p_{3/2}}}{60},
\end{equation}
\end{small}
\begin{small}
\begin{equation}
\label{eq2}\delta W_{2}=2A^{7p_{3/2}}-B^{7p_{3/2}}-28C^{7p_{3/2}}+\frac{\eta^{7p_{3/2}}}{45}+\frac{2\sqrt{5} \zeta^{7p_{3/2}}}{75},
\end{equation}
\end{small}
\begin{small}
\begin{equation}
\label{eq3}\delta W_{3}=3A^{7p_{3/2}}+B^{7p_{3/2}}+8C^{7p_{3/2}}-\frac{\eta^{7p_{3/2}}}{20}-\frac{\sqrt{5} \zeta^{7p_{3/2}}}{100}.
\end{equation}
\end{small}
Setting Eqs.(\ref{eq1})$\times$5-Eqs.(\ref{eq2})$\times$4+Eqs.(\ref{eq3}), one can find
\begin{small}
\begin{equation}
C^{7p_{3/2}}=\frac{1}{400}\left(5\delta W_{1}-4\delta W_{2}+\delta W_{3} +\frac{\sqrt{5}\zeta^{7p_{3/2}}}{5}\right).
\end{equation}
\end{small}

It can be deduced that the sensitivity of measurements on these hyperfine splittings on the order of $\sigma_{\delta W} \approx 1\mathrm{kHz}$ would result in an uncertainty in the HFS constant $C$ on the order of $\sigma_{C} \approx 0.0162 \sigma_{\delta W} \approx 20 \mathrm{Hz}$, and the effects of HFS constant $C$ of the predicted magnitude would be revealed within this sensitively reliably.

It is also important to note that $C$ is related to $\zeta^{7p_{3/2}}$ factor, which represents the second-order correction. If we want to evaluate $\zeta^{7p_{3/2}}$, we have to know the off-diagonal elements which are also one of the major systematics in the extractions of the $C$ values from the measured hyperfine splittings. For this purpose, we have calculated these matrix elements using the DF, LCCSD and CCSD methods and displayed them in Table \ref{6}. Then we find $\eta^{7p_{3/2}}$=0.051$\mathrm{KHz}$, $\zeta^{7p_{3/2}}$=7.09$\mathrm{KHz}$, and
\begin{small}
\begin{equation}
\Delta C^{7p_{3/2}}=\frac{\sqrt{5}\zeta^{7p_{3/2}}}{2000}=7.93 \mathrm{Hz}.
\end{equation}
\end{small}
For the $\zeta^{7p_{3/2}}$ parameter, the contribution from off-diagonal matrix elements between $7p_{3/2}$ and $7p_{1/2}$ states with the lowest denominator of the energy difference is predominant, and the rest can be ignored.
\begin{table}[h]\small
\begin{threeparttable}
\newcommand{\RNum}[1]{\uppercase\expandafter{\romannumeral #1\relax}}
\caption{  Important off-diagonal matrix elements (in MHz) among the fine-structure partners obtained using our DF, LCCSD and CCSD  methods.}\label{6}
\begin{ruledtabular}
\begin {tabular}{cccccc}
\toprule
$\text {Off-diagonal matrix}$ &$\text{DF}$ &$\text{LCCSD}$  &$\rm CCSD_{L}$ &CCSD&Final    \\
\hline
$\langle 7p_{1/2}||{O}^{(1)}|| 7p_{3/2}\rangle$   &$-$229      &$-$147          &$-$123  &$-$119   &$-$119(5)      \\
$\langle 6d_{3/2}||{O}^{(1)}|| 6d_{5/2}\rangle$   &$-$125.5    &$-$2536         &$-$2350 &$-$2449  &$-$2449(116)    \\
$\langle 7p_{1/2}|| O^{(2)}||  7p_{3/2}\rangle$   &$-$1281     &$-$2222         &$-$2155 &$-$2158  &$-$2158(44)       \\
$\langle 6d_{3/2}|| O^{(2)}||  6d_{5/2}\rangle$   &$-$245      &$-$334          &$-$338  &$-$334   &$-$334(5)       \\
\end{tabular}
\end{ruledtabular}
\end{threeparttable}
\end{table}

\section{Conclusion}
In summary, we calculated energy levels and hyperfine-structure coupling constants of Ra$^{+}$ using Dirac-Fock (DF), lower-order many-body perturbation theory (MBPT), coupled-cluster single- and double-excitation approximation with (CCSD) and without non-linear terms (LCCSD).
The calculated energies and HFS constants $A$ show a good agreement with available experimental values. We find that for high angular momentum states, such as $6d_{5/2}$ state, the BW effect is very important, and this contribution mainly comes from the calculation of the correlation effect.
To more accurately calculate the contribution of BW to the hyperfine structure constant of HFS constant $A$ in the low energy state of radium ions, further optimization of the model is needed, such as considering the wave function of the nucleus and the spin interaction, etc. Combining with available experimental HFS constants $B$, we evaluated the electric quadrupole moments $Q$ for $^{209,211,221,223}$Ra.
For $^{221,223}$Ra isotopes, our $Q$ values are consistent with the values from~\cite{Neu1988epd} that are widely referenced today within $0.5\%$. We recommended a value of $Q=0.33(2)~b$ for $^{211}$Ra nucleus, which is smaller than the referenced value of 0.48(4)~b about 30\%. If HFS constants $B$ of other isotopes are obtained, corresponding $Q$ can be also extracted by combining our recommended results. In addition, considering the preliminary value of $\Omega$ from the nuclear shell model, its contributions to the hyperfine structures of some low-lying states in Ra$^{+}$ are estimated. We conclude that it would be capable of revealing the effects of the HFS constant $C$ of the present magnitude of experimental measurement. Furthermore, the expressions for the hyperfine splitting for some important low-lying states derived in the present work are useful for extracting out the HFS constants $A$, $B$, and $C$ when these hyperfine-structure levels in Ra$^{+}$ is measured with high-precision.

\begin{acknowledgments}
We are grateful to Yong-jun Cheng of Shaanxi Normal University and Ji-Guang Li of Institute of Applied Physics and Computational Mathematics for
many useful discussions. The work was supported by the National Natural Science Foundation of China (No. 11504094, No. 11674253, No. 11774386), the Strategic Priority Research Program of the Chinese Academy of Sciences Grant (No. XDB21030300), the National Key Research and Development Program of China under Grant (No. 2017YFA0304402), the Post-doctoral research project of SZTU (No. 202028555301011), and the project of Educational Commission of Guangdong Province of China (No.2020KTSCX124).

\end{acknowledgments}

\section*{}
%

\end{document}